%% file: sample-manuscript.tex
\documentclass[acmlarge]{acmart}

\AtBeginDocument{%
  }
\setcopyright{cc}
\setcctype{by}
\acmJournal{IMWUT}
\acmYear{2025} \acmVolume{9} \acmNumber{4} \acmArticle{178} \acmMonth{12} \acmDOI{10.1145/3770654}

\newcommand{\DataTap}{TapNav}

\usepackage{array}

\begin{document}

\title{TapNav: Adaptive Spatiotactile Screen Readers for Tactually Guided Touchscreen Interactions for Blind and Low Vision People}

\author{Ricardo E. Gonzalez Penuela}
\affiliation{%
  \institution{Cornell University}
  \city{New York}
  \country{USA}}
\email{reg258@cornell.edu}
\affiliation{%
  \institution{JPMorgan Chase}
  \city{New York}
  \country{USA}}

\author{Fannie Liu}
\affiliation{%
  \institution{JPMorgan Chase}
  \city{New York}
  \country{USA}}
\email{fannie.liu@jpmchase.com}

\author{Blair MacIntyre}
\affiliation{%
  \institution{JPMorgan Chase}
  \city{New York}
  \country{USA}}
\email{blair.macintyre@jpmchase.com}

\author{David Saffo}
\affiliation{%
  \institution{JPMorgan Chase}
  \city{New York}
  \country{USA}}
\email{david.saffo@jpmchase.com}

\renewcommand{\shortauthors}{Gonzalez et al.}

\begin{abstract}
Screen readers are audio-based software that Blind and Low Vision (BLV) people use to interact with computing devices, such as tablets and smartphones. Although this technology has significantly improved the accessibility of touchscreen devices, the sequential nature of audio limits the bandwidth of information users can receive and process. We introduce TapNav, an \textit{adaptive spatiotactile screen reader} prototype developed to interact with touchscreen interfaces spatially. TapNav's screen reader provides adaptive auditory feedback that, in combination with a tactile overlay, conveys spatial information and location of interface elements on-screen. We evaluated TapNav with 12 BLV users who interacted with TapNav to explore a data visualization and interact with a bank transactions application. Our qualitative findings show that touch points and spatially constrained navigation helped users anticipate outcomes for faster exploration, and offload cognitive load to touch. We provide design guidelines for creating tactile overlays for \textit{adaptive spatiotactile screen readers} and discuss their generalizability beyond our exploratory data analysis and everyday application navigation scenarios.
\end{abstract}



\begin{teaserfigure}
    \centering
    \includegraphics[width=0.9\linewidth]{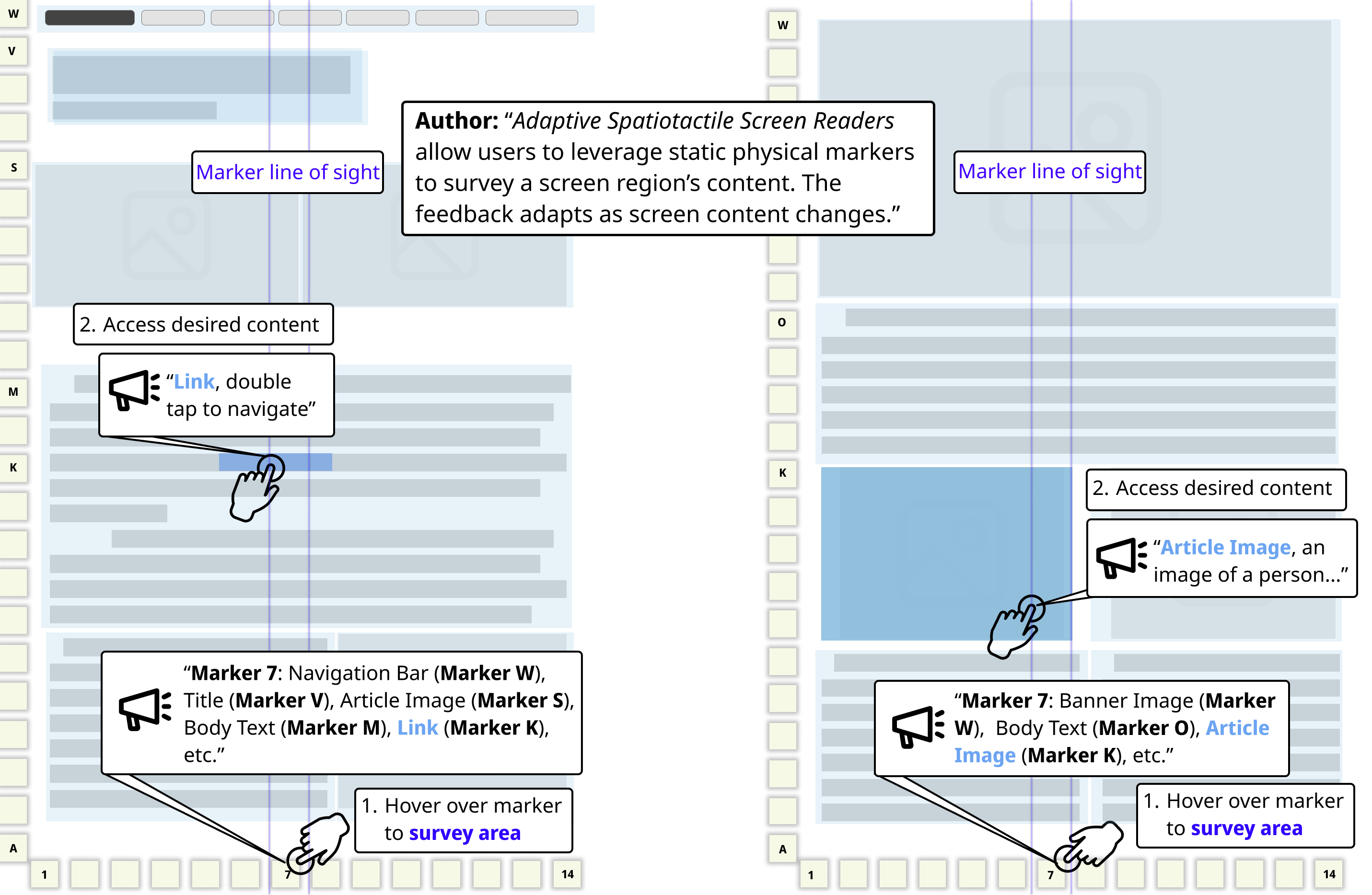}
    \caption{ TapNav, our \textit{adaptive spatiotactile screen reader}. It uses physical markers (green) on a tactile overlay layered over a touchscreen device to serve as anchor points and scaffold spatial interactions. Blind and Low Vision users can touch these markers, and the screen reader adapts its feedback as the on-screen content changes (blue). Here, triggering marker 7 yields two different screen reader outputs, informing the user of the current content within marker 7's "line of sight".}
    \Description{TapNav, our adaptive spatiotactile screen reader. It uses physical markers  on a tactile overlay layered over a touchscreen device to serve as anchor points and scaffold spatial interactions. Blind and Low Vision users can touch these markers, and the screen reader adapts its feedback as the on-screen content changes. Here, triggering marker 7 yields two different screen reader outputs, informing the user of the current content within marker 7's "line of sight".}
    \label{fig:teaser}
\end{teaserfigure}

\maketitle

\input{src/01_introduction}

\input{src/02_related_work}

\input{src/03_Design_Goals_Rationale}

\input{src/05_system}
\input{src/06_evaluation}

\input{src/07_findings}
\input{src/08_discussion}
\input{src/09_conclusion}

\section{Acknowledgments}
We would like to express our appreciation to the many teams at JPMorgan Chase Bank, N.A. for their support and engagement throughout this project. Their involvement and collaboration contributed to the development and completion of this work. We would also like to thank the Global Technology Applied Research center of JPmorgan Chase researchers for their feedback throughout the execution of this project.

\section{Disclaimer}
This paper was prepared for informational purposes by the Global Technology Applied Research center of JPMorgan Chase. This paper is not a product of the Research Department of JPMorgan Chase or its affiliates. Neither JPMorgan Chase nor any of its affiliates makes any explicit or implied representation or warranty and none of them accept any liability in connection with this paper, including, without limitation, with respect to the completeness, accuracy, or reliability of the information contained herein and the potential legal, compliance, tax, or accounting effects thereof. This document is not intended as investment research or investment advice, or as a recommendation, offer, or solicitation for the purchase or sale of any security, financial instrument, financial product or service, or to be used in any way for evaluating the merits of participating in any transaction.
\bibliographystyle{ACM-Reference-Format}
\bibliography{sample-base}
\end{document}

%% file: src/01_introduction.tex
\section{Introduction}

Touchscreen devices are commonly used for communication, accessing online services, and productivity. However, these devices predominantly rely on visual cues to convey information, creating significant barriers for Blind and Low Vision (BLV) people. To compensate, screen readers enable BLV people to navigate touchscreen devices by detecting touch gestures and translating interface elements into spoken descriptions; however, this is not without limitations. First, screen readers convey information sequentially—users must listen to each screen element’s description before moving to the next—which caps interface navigation speed to the speech rate users can comprehend. Second, users must mentally reconstruct spatial relationships and remember previously heard information while they actively navigate auditorily,  increasing cognitive load considerably \cite{brown2003design}.

For example, imagine a BLV renter is exploring a scatter plot of apartment size versus rent in New York City. The screen reader speaks ``Manhattan, 33rd St, 500 ft², \$3,200''; ``Brooklyn, 55 Clark St, 650 ft², \$2,100''; and so on for multiple points. Although each data point is accessible, linear narration makes it challenging to perceive patterns like clusters of affordable, larger apartments. Similar challenges emerge in everyday applications with table structures \cite{jiang2023understanding}, such as navigating banking transactions. In these applications users must traverse entries sequentially, making it difficult to identify spending patterns. In short, audio alone conveys values but not the spatial relationships for fully understanding data representations \cite{munzner2014visualization, jung2021communicating}, exploration, and interface navigation that rely on position as a means to encode information and define interaction \cite{munzner2014visualization}.

To address this, researchers have explored two types of cognitive scaffolds to help BLV users form spatial mental models of on screen content: audio-based and touch-based scaffolds. Audio-centric research explores how to leverage the auditory channel by mapping values to pitch, rhythm, or timbre to ``sonify'' graphical information (such as data visualizations) \cite{madhyastha1995data, wang2022seeing,hoque2023accessible} or using modulated spatialized audio cues that encode position and proximity \cite{bru2023line, ahmetovic2021multi, zhao2008data, holloway2022infosonics, zhao2024tada}. Although these techniques can help users explore data representations and approximate the positions of interface elements, they still require users to reconstruct complex layouts mentally and heavily rely on memorization rather than recognition.

In contrast, touch-centric methods shift the cognitive burden from auditory memory to tactile recognition by pairing screen readers and audio cues with physical feedback. Researchers have explored several methods for augmenting touchscreen screen readers with physical feedback: embossed or removable overlays \cite{kane2013touchplates, el2013touch, avila2018tactile, mackowski2023multimodal,melfi2020understanding, he2017tactile}, vibrotactile output \cite{el2018evaluating, jiang2024designing, soviak2015feel, kim2016assisting}, and constrained interaction techniques that limit user actions to specific touch regions \cite{chundury2023tactualplot, nikitenko2014touching, mackowski2023alternative, zhao2024ai, zong2024umwelt}. By leveraging proprioception and tactile perception, these methods provide tactile reference points that, together with screen readers and audio cues, favor recognition over memorization \cite{palivcova2020interactive}. We refer to these systems as \textit{spatiotactile screen readers}, a subset of screen readers that integrate spatially organized tactile feedback (e.g., physical markers or overlays) with auditory output. Unlike mainstream touch-based screen readers like VoiceOver and Talkback, \textit{spatiotactile screen readers} are explicitly designed to bind physical cues on a surface to specific digital elements (e.g., touching Braille marker "3/C" triggers the feedback ``marker 3 contains four high-rated movies'').

Despite their promise, \textit{spatiotactile screen readers} remain narrowly scoped. Tactile overlays used in \textit{spatiotactile screen readers} are usually handcrafted or embossed for a single, static graphic \cite{avila2018tactile}. Thus, when the underlying interface changes or when users switch to a different app, there is a mismatch between the tactile overlay and the content rendered. We believe there is an opportunity to design \textit{spatiotactile screen readers} that can adapt their feedback to interface changes on the fly while continuing to scaffold their guidance with direct references to tactile overlays cues (e.g., markers, landmarks, etc.) designed for general use. 

However, designing \textit{spatiotactile screen readers} that can adapt across use cases presents several challenges. First, a tactile overlay must accommodate different layouts while maintaining consistent spatial mappings. Second, the system feedback must be designed in such a way that it meaningfully binds the tactile cues to digital content without overwhelming the user. Third, designers must balance simplicity and specificity needed for effective task support (e.g., accommodate general-purpose versus task specific).
 
To explore these considerations, we introduce \textbf{adaptive} \textit{spatiotactile screen readers} to support diverse everyday tasks. To this end, we developed and evaluated \DataTap{}, a grid based \textit{adaptive spatiotactile screen reader}. \DataTap{} delivers concise, context dependent verbal prompts (e.g., ``four movies with 8.5 critic rating'', ``5 screen elements on row 10, selecting first'') and can constrain subsequent user input gestures to selected regions, enabling rapid, spatially focused and tactually supported exploration (See Fig. \ref{fig:teaser}).

Using \DataTap{}, we conducted a user study with 12 BLV participants and evaluated it in two representative scenarios across different tasks and context: (1) exploratory data analysis—discovering information and locating data points in a scatter plot that compares review scores of movies from different populations; and (2) everyday application navigation—navigating a bank account transactions in a mobile banking interface with a table structure. 

Through the design and evaluation of \DataTap{}, an exploratory probe for adaptive \textit{spatiotactile screen readers}, our work provides the following contributions:

\begin{itemize}
    \item We provide insights from Blind and Low Vision people highlighting how touch points and spatially constrained navigation can improve exploration efficiency by shifting some auditory information to tactile perception. This approach helps  BLV users anticipate interaction outcomes (e.g., ``The first marker should give me information about the movies with lowest ratings'') to reduce cognitive load and increase navigation speed.
    
    \item We identify three design levers for \textit{adaptive spatiotactile screen readers} to support access and spatial understanding of on-screen content: orientation cues to support localization and alignment, marker styles matched to users abilities and Braille literacy (e.g., braille vs non-braille, embossed vs cutouts), and alternative screen reader navigable structures to support sensemaking.       
\end{itemize}

We conclude by providing design guidelines for creating tactile overlays, discuss the generalizability of \textit{adaptive spatiotactile screen readers} and their limitations, and how our approach could be extended to broader application domains beyond our exploratory data analysis and everyday application navigation scenarios.

%% file: src/02_related_work.tex
\section{Related Work}

Our work contributes to research on approaches to improve touchscreen accessibility for BLV people, particularly for spatial understanding of interface layout. We draw from two major threads: audio-based approaches to convey information of spatially complex visual content (like data visualizations), and touch-based approaches that leverage physical feedback to support spatial understanding.

\subsection{Audio-based Approaches for Touchscreen Accessibility}

Screen readers such as VoiceOver\cite{apple_voiceover} and TalkBack \cite{google_talkback} provide BLV people access to touchscreen devices: they map touch gestures to spoken feedback and describe the underlying UI elements on screen, enabling BLV people explore and activate controls sequentially. 

However, this linear narration ``flattens'' the interface, losing both spatial layout and hierarchical spatial relationships of screen elements that sighted people can perceive at a glance. For BLV people, accessing spatially separated points on a screen is difficult because touchscreen devices' smooth surface lacks any differentiable tactile qualities \cite{kane2008slide}: a point close to a corner feels the same way as a point in the middle of the screen. Thus, while navigating an interface, BLV people are forced to hold lists of screen elements in working memory and to reconstruct spatial relationships mentally \cite{wijntjes2008look}.  This cognitive burden is exacerbated when using applications with interfaces designed primarily for sighted people \cite{mcgookin2008investigating,kane2008slide}. To mitigate these shortcomings, researchers have explored enhancing the auditory channel in two main ways.

\subsubsection{Structured Screen Reader Audio}

To ease the burden of sequential, element-by-element exploration, researchers have added structured audio layers that break the screen into higher-level regions \cite{kane2008slide,kane2011access,li2022enhancing}, or provide auditory menus \cite{yalla2008advanced}. Early work such as ``Slide Rule'' remapped swipe gestures to move a ``focus cursor'' around the interface; although still linear, it let users jump row‑by‑row rather than pixel‑by‑pixel. Building on that principle, Access Overlays defined regions along a screen’s edges—gliding a finger along one of these regions announces the section’s label, shortening target‑acquisition time on large displays \cite{kane2011access}. Hierarchical auditory menus like \citet{yalla2008advanced}’s framework exposes nested UI structures through consistent shortcut gestures, allowing users to skip entire branches of content \cite{yalla2008advanced}.

\subsubsection{Spatialized Audio Cues for Graphical Information Accessibility}
Spatialized audio cues address two needs: making graphical outlines and regions audible and conveying data-dependent properties. When the task is inherently spatial—tracing roads, reading maps, or  node diagrams—researchers add position-dependent audio cues based on users' touch \cite{delogu2010non,cohen2006teaching,zhao2024tada}. As the user slides a finger, parameters such as pitch or timbre change continuously, turning edges and regions into audible contours \cite{kennel1996audiograf,yoshida2011edgesonic,poppinga2011touchover}. For example, TADA lets BLV users follow links in dense network diagrams by generating auditory cues through the paths of the nodes in the diagram \cite{zhao2024tada}. Similarly, \citet{ahmetovic2021multi} map pitch to a finger’s location along a line segment, letting listeners sense its orientation and length in real time.

When the goal is to interpret dense graphical information—such as data visualizations, where users must identify chart trends, compare categories, or spot clusters—researchers attach data-dependent audio cues directly to the graphics \cite{madhyastha1995data,zhao2008data}. As the listener navigates through the visualization, parameters such as pitch, or rhythm represent data magnitude and type, turning points and series into audible patterns \cite{bru2023line,chundury2023tactualplot}. For example, Line Harp sweeps pitch along the x-axis so rising melodies reveal upward trends when an user is reading a line chart.

\subsubsection{Limitations of Audio-based Approaches}
While audio menus, spatialized cues, and sonification have made touchscreen content more accessible, these approaches have limitations. Audio interfaces can only convey small amounts of spatial information at once, requiring  sequences of actions for both exploration and input \cite{palani2017principles}. This creates a load on users' working memory \cite{brown2003design}, and even with accessibility aids, BLV users' navigation with screen readers remains slow and inaccurate with complex graphical information \cite{sharif2021understanding}

Behavioral and neuro-imaging evidence positions touch as the closest non-visual analogue to vision for spatial reasoning—and, unlike sound, it does not obfuscate other simultaneous cues \cite{giudice2018navigating, giudice2011functional}. In the context of complex graphical information, \citet{wall2006feeling} suggest that haptics should convey shapes and regions, while audio should be used to convey fine-grained details. Thus, researchers have explored approaches that incorporate physical feedback for touchscreen accessibility.

\subsection{Touch-based Approaches for Touchscreen Accessibility}

Touch can convey information such as shapes, boundaries, and spatial configurations in parallel with audio, easing the working-memory demands that arise with audio-only solutions. To leverage this strength on touchscreens, researchers have explored three main strategies: (1) physicalizing UI controls through hardware add-ons, (2) creating tactile overlays that augment screen readers, and (3) using vibrotactile feedback to mark positions and screen elements. We refer to approaches 2 and 3 as \textit{spatiotactile screen readers}—systems that provide physical feedback through tactile overlays or vibrotactile output, where each physical element corresponds to specific digital content.  We review these three strategies in the following subsections.

\subsubsection{UI Physicalization With Hardware} \label{subsubsection:physicalization}

To provide BLV users with tactile, task-specific controls, researchers have explored physicalizing interface UI controllers as external hardware \cite{li2022tangiblegrid, li2019editing, zhang2018interactiles} and alternative input devices \cite{islam2024wheeler,stearns2018touchcam}.  This idea has been widely explored in desktop environments in the context of providing access to webpage layouts, specifically to support BLV users design their own web layouts \cite{li2019editing,li2022tangiblegrid, potluri2019multi}. For example, TangibleGrid maps webpage regions to snap-in, shape-changing brackets on a sensorized board, giving Blind designers real-time tactile and spoken feedback as they arrange blocks. In a user study, TangibleGrid supported participants to understand and create layouts non-visually, showing how physical UI controllers can externalize page structure \cite{li2022tangiblegrid}. In the context of touchscreen devices, researchers have created systems like Interactiles, a low-cost toolkit that augments smartphones with snap-on 3-D-printed keypads and scroll wheels \cite{zhang2018interactiles}. In a user study with Interactiles, researchers found that participants navigated faster than when using gesture-only screen-reader interaction. In short, hardware add-ons reveal structure and accelerate specific tasks by physicalizing the controls designers anticipate and users must carry the extra hardware with them.

\subsubsection{Single-Layout Tactile Overlays}

Raised-line, cutouts and textured sheets have long enabled BLV users to explore maps, diagrams, and other spatial content through touch \cite{wall2006feeling}. To improve touchscreen accessibility, this idea was implemented by overlaying fixed plates that align with on-screen elements \cite{kane2013touchplates, soviak2015feel}. For example, Touchplates \cite{kane2013touchplates} let users overlay plastic cutouts of physical keyboards, numpads, or map guides onto a large display and reported faster target-acquisition times and higher confidence than audio-only exploration. Other researchers designed one-to-one overlays to help users understand other kinds of layouts such as document structure \cite{avila2018tactile}, STEM education diagrams \cite{melfi2020understanding, mackowski2023multimodal} and floor-plans \cite{he2017tactile}. 

\subsubsection{Vibrotactile Feedback}
Researchers have also explored encoding spatial cues through the vibration motors built into touchscreen devices \cite{zhang2024charta11y,el2018evaluating} or through external aids \cite{kim2016assisting,jiang2024designing}. Brief pulses aligned with interface elements let BLV users trace lines and locate targets; because each pulse is tied to a screen position, it simulates “touching” the content and supports mental mapping without extra verbal load. For example, in ChartA11y these location-specific cues helped participants detect scatter-plot trends more accurately than audio-only approaches \cite{zhang2024charta11y}. On tablets, \citet{el2018evaluating} show that vibrotactile targets obey Fitts’ Law for Blind users under typical size and distance parameters, suggesting predictable performance bounds for vibration-based pointing \cite{el2018evaluating}.

\subsubsection{Limitations of Touch-based Approaches}

To summarize then, current touch-based solutions overcome some of the issues with audio-only approaches but face a trade-off between rich feedback and adaptability. Hardware add-ons are less portable and have not scaled outside research prototypes. Although effective at conveying spatial information, overlays must be fabricated for each screen state, so the tactile cues break whenever the content or application changes. As for vibrotactile feedback, these cues share similar transient and low-resolution properties as audio, offering few persistent landmarks as content changes. To address these limitations, our work explores \textit{adaptive spatiotactile screen readers} that combine persistent tactile landmarks with context-dependent speech prompts and audio cues to adapts across interfaces and application changes.

%% file: src/03_Design_Goals_Rationale.tex
\section{Formative Studies}

To ensure our work was adequately grounded in real needs and use cases for BLV people, we conducted two formative studies. First, a pilot test to inform the feasibility and usability of combining tactile overlays with  spatial screen reader interactions for touch screen content (\textit{adaptive spatiotactile screen readers}). Second, a focus group study with professionals who have experience working with BLV people to complement the technical insights from the pilot with real-world application scenarios for our main study.

\subsection{Pilot Study} \label{subsection:pilot}
To inform our initial hypothesis of the utility of \textit{adaptive spatiotactile screen readers}, we conducted a pilot test with an initial proof of concept version of \DataTap{}. 
The goal of this test was twofold:
\begin{itemize}
    \item Probe the conceptual utility of \textit{adaptive spatiotactile screen readers} with BLV people and accessibility experts to ensure grounding and community participation. 
    \item Test and explore the design of tactile overlays, touch gestures, and interactions to inform usability design moving forward. 
\end{itemize}

The \DataTap{} proof of concept was created to present a simple scatter plot visualization that the user could explore spatially. This scenario was chosen because data visualizations, such as scatter plots, rely heavily on spatial encoding, are highly dynamic as they change with the data and scales being encoded, and often depend on user interactions such as accessing details on demand, filtering, and zooming \cite{munzner2014visualization}.

This scenario featured a simple overlay that acted as an axis for the visualization, featuring repeating cutout shapes to indicate significant tick marks, and lines to indicate axis sections.
 We based our gestures design on Access Overlays \cite{kane2011access} ``Edge Projection'', and VoiceOver's touchscreen gestures  \cite{apple_voiceover}.
Users could interact with markers to receive screen reader feedback about the points located in line-of-sight of the selected tick mark (similar to ``Edge Projection'', see figure \ref{fig:dataviz}). 
Users could also move along axis lines to locate and tap individual data points to receive screen reader prompts for their details (Similar to VoiceOver). 
More details about this can be found in Section \ref{sec:system_top_level}.
To determine the size of the markers of the overlay, we conducted an internal testing session within our research group where we evaluated which size was best to reliably perceive and differentiate markers (See Fig. \ref{fig:markerdesign}).

Using this proof of concept, we conducted usability tests with two members of our organization, one self-reporting as Blind, who was a technology professional and screen reader user, and one sighted person who was an expert in screen reader and digital accessibility technologies (Both with experience interacting with data visualizations in professional settings and reading alternative text descriptions for data visualizations). 
Sessions began with an overview of the components of the proof of concept of \DataTap{}, including the vertical and horizontal axes, physical markers, and data points. Participants practiced tapping and dragging gestures to receive auditory feedback about the data.
They explored the iris dataset to familiarize themselves with the prototype's functionality before moving on to a dataset of 36 movie ratings.
Participants completed tasks to assess their ability to navigate and extract information, including search tasks to locate specific movies, query tasks to identify movies with certain attributes, and exploratory tasks to find movies of personal interest.

After completing the tasks, we performed a semi-structured interview to gather feedback on their experience with the prototype. 
The interview focused on ease of use, likes and dislikes, suggestions for improvement for the components of the proof of concept, and feedback on the recruitment process and study accessibility. 
Data was collected through audio recordings of the sessions, and researcher notes, capturing both participant interactions and verbal feedback about the experience completing the tasks with the prototype.

By analyzing this feedback, we generated insights regarding the conceptual viability of \textit{adaptive spatiotactile screen readers} and practical design challenges encountered during the pilot:
 \begin{itemize}
     \item Participants envisioned \DataTap{}'s proof of concept as a tool to support general-purpose interface navigation access. Although the proof of concept was introduced to participants as a system primarily for accessing data visualizations, both participants unprompted mentioned its utility went beyond data visualizations and would use it in other kinds of applications. Our Blind participant mentioned they ``wished'' there was ``some tool like this to help access material during college, especially for STEM degrees''.
     \item For the search task and the query task, participants found the tactile markers useful  for browsing data by touch (e.g., ``let me check the movies with highest critic reviews''). The physical feedback allowed them to revisit screen regions reliably, and thus compare information or access details on demand.
     \item Participants had difficulty performing straight-line dragging gestures. While the tactile markers supported localizing some screen regions reliably, we observed that participants struggled to pinpoint data points located in the middle of the screen because their finger would veer off course. 
     \item Participants found the cutouts markers size appropriate (0.75 cm by 0.75 cm) but also recommended exploring other designs as some of the cutouts were sometimes difficult to differentiate (e.g., is this a square or a pentagon?).
 \end{itemize}

These insights addressed our pilot goals. First, we confirmed the conceptual utility of \textit{adaptive spatiotactile screen readers}, as participants found the proof of concept useful for data visualizations analysis and also envisioned applications beyond our initial scope, suggesting broader potential for accessing touchscreen devices. Second, we identified specific challenges with gesture interaction and markers perception, providing clear direction for design improvements. 

As a result, we enhanced our proof of concept with several new features and explored additional designs for the main user study. For example, we developed the spatial navigation mode in the interface navigation prototype (See Section \ref{subsec:spatialnavigation}) to support quick location-based access to screen elements positioned in the middle of the screen. We also introduced Braille raised markers as an alternative design to the cutout markers to leverage BLV users' prior experience with this medium (See Section \ref{subsubsec:tactile_overlays_braille}).

\subsection{Focus Groups}
\label{subsection:focus}

To explore real world scenarios and use cases for \textit{adaptive spatiotactile screen readers}, we conducted workshops with bank employees at different managerial levels of responsibility who perform customer-facing job functions, and whose services are often accessed through digital interfaces.
We had \textbf{two primary goals} for these workshops: 
\begin{itemize}
    \item Gather insights on current practices for interacting with BLV people, especially as it relates to interacting with digital tools and services. 
    \item Brainstorm use cases where \DataTap{} could enhance accessibility to better serve BLV people within these job functions, in order to develop realistic scenarios for evaluation.  
\end{itemize}

We held three workshops, each with different sets of 4-5 participants, representing various job functions (including bank associates, lead associates, and branch managers) and geographical regions within our organization. This diversity provided a comprehensive view of customer interactions across different contexts. No BLV people were involved in the workshops. Due to data privacy protection policies for employees within our company, we did not collect additional demographic data. However, given their professional roles involving customer-facing digital services support, participants had practical experience with the types of interfaces and workflows relevant to our study.

Sessions comprised of 60-minute virtual workshops using a shared digital collaboration board. 
After an introduction, participants listed and described their main customer-facing functions, categorizing them as they were mentioned. 
This was followed by a discussion on the current state of accessibility for BLV users engaging with our digital services, where participants shared their experiences, with notes recorded on the digital board.
Then, the proof of concept of \DataTap{}  was introduced as a potential assistive technology solution to help BLV people navigate touchscreen devices. Participants brainstormed scenarios for integrating \DataTap{} as a tool to support customer-facing functions and created storyboards envisioning how a BLV person could use \DataTap{} to complete a task.
Groups then discussed these storyboards, providing feedback on \DataTap{}'s applicability in enhancing accessibility.

Data was collected through audio recordings, digital board notes, and participant-generated storyboards, offering qualitative insights into accessibility challenges and potential use scenarios for \DataTap{}. One researcher organized digital board notes and participants quotes through one round of affinity diagramming to generate codes and group them into themes \cite{hartson2012ux}. Then, three researchers discussed and validated these groupings. This thematic analysis revealed \textbf{three primary findings}:

\begin{itemize}
    \item BLV people face challenges accessing digital services independently. Some BLV people often come to meet agents in person for help because they struggle to find features, settings, and other business services provided through mobile applications (e.g., changing password, updating personal address, etc.).
    
    \item Customer-facing agents, despite willing to provide support, are prohibited from directly interacting with BLV people's devices. Based on privacy and industry policy, even if agents want to directly assist BLV people with their needs, customers are required to perform actions and inputs on their own. Currently, as a best alternative, agents "mediate" access to visual information by answering questions and providing verbal guidance ad hoc.

    \item Customer-facing agents face challenges when mediating access to visual information and providing guidance to BLV people using accessibility tools. While agents can describe what appears on screen or answer questions about digital documents, the mismatch between their experience using touchscreens and BLV people's makes it difficult for agents to give navigation instructions that align with how BLV people actually experience digital interfaces when using their own devices (e.g., an agent might say ``click the red button in the top right corner'' when the screen reader announces it as ``submit button, link'' in a different reading order).
\end{itemize}

Based on these findings, we addressed our focus group goals as follows: First, we gathered insights on current practices for supporting BLV customers with digital services access in-person, and we learned that bank agents serve as mediators who provide verbal guidance but face challenges due to policy constraints and mismatches between visual and screen reader navigation. These insights validated the need for \DataTap{}'s spatial navigation approach in real-world service contexts.  Second, we brainstormed use cases and incorporated two identified scenarios into our main study: one for the familiarization section where users would navigate a PDF document with complex layout elements (tables and dense textual information), and another for the evaluation section that was representative of applications common in daily life. Specifically, we chose to recreate part of a bank transactions application featuring typical GUI elements (e.g., buttons, navbars, links, etc.). These scenarios ensured our main study covered visually dense tasks (data visualization scenario) and high-frequency interfaces that BLV users commonly encounter with barriers of access.

%% file: src/05_system.tex
\section{Design of TapNav} \label{sec:system_top_level}

\DataTap{} is a multimodal system that combines tactile overlays, auditory feedback, and spatially constrained touchscreen interactions to create accessible representations of traditionally visually oriented interfaces. Through a combination of exploration by touch and screen reader gestures, BLV users can navigate interfaces efficiently while learning the spatial arrangement of interface components. \DataTap{} layers tactile overlays on touchscreen devices to provide physical landmarks that users touch to guide their interactions. Using the tactile overlays as reference points, \DataTap{} enables spatially grounded interactions that allow BLV users to contextualize their interface exploration. While \DataTap{} represents a spatial-tactile-first approach, meaning the goal is for users to understand spatial arrangement leveraging the sense of touch, we implemented traditional navigation methods from screen reader technology (like swiping gestures and others) to complement spatial navigation.

In the following section, we present a brief overview of the design rationale for each core component of \DataTap{} and the screen reader experience (gestures, feedback, and audio cues) we implemented for two prototypes that use \DataTap{}, a data visualization prototype and an interface navigation prototype.

\subsection{Tactile Overlays} \label{subsec:tactile_overlays}

\input{tables/tactile}
 \begin{figure}
    \centering
    \includegraphics[width=0.65\linewidth]{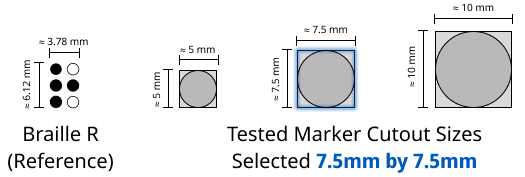}
    \caption{The size of the braille and shape markers tested and used (outlined in blue) for the tactile overlays.}
    \label{fig:markerdesign}
    \Description{On the left side, an image of a Braille letter "r" showing size dimensions 3.78mm width by 6.12mm height. This was the approximate size of the braille markers used for the interface navigation prototype. On the right side, three reference images of "circle" cutout markers of different sizings tested ( 5mm by 5mm, 7.5mm by 7.5mm, and 10mm by 10mm). The 7.5mm by 7.5mm marker is highlighted indicating that this size was the final size selected for the data visualization prototype.}
\end{figure}

Tactile overlays provide a tangible reference structure that users can perceive through touch, anchoring their navigation, selection, and interaction with interface elements. 
We created two overlays to explore user preferences for two of the most representative methods for producing tactile graphics: raised Braille, and cutout outlines. The first leverages users' prior knowledge of Braille (Braille Raised Markers), and the second requires minimal prior knowledge (Geometric Shape Cutout Markers).
Both overlays were designed following a similar principle: The screen would be divided into rows and columns. There would be an ``X-axis'' and a ``Y-axis'' for both overlays. Many researchers have found that this arrangement is one of the most effective when dividing content logically for accessible data visualizations to BLV users \cite{zhang2024charta11y, chundury2023tactualplot}. Each marker on the axis represents either a row or a column. While it is possible to create other arrangements of screen divisions, it is out of the scope of this exploratory study. Finally, both overlays measured 26.7 cm by 16.7 cm based on the touchscreen area of the Samsung Galaxy Tab S8+, the device we used to conduct our study. 

\begin{figure}
    \centering
    \includegraphics[width=0.85\linewidth]{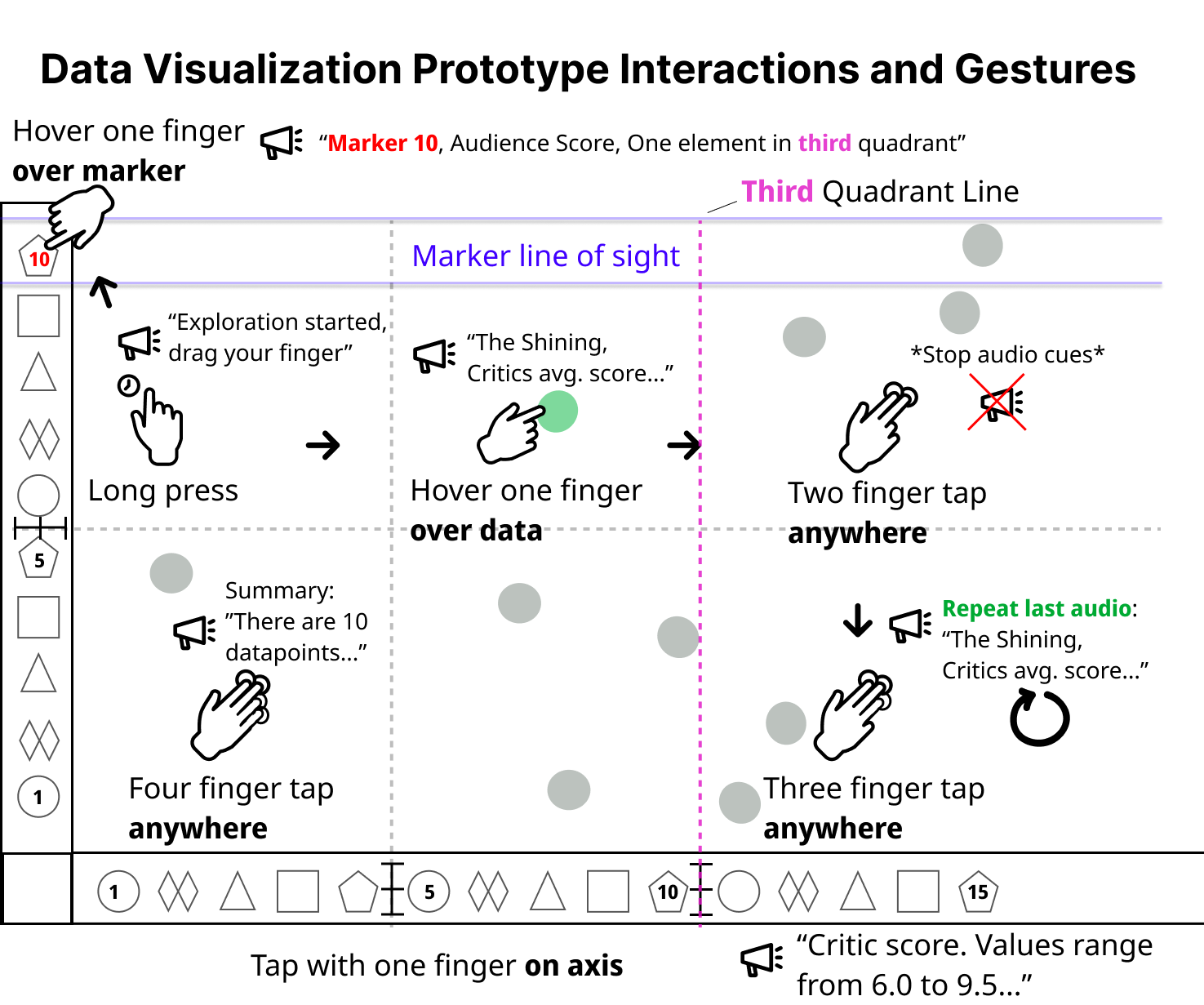}
    \caption{Participants can start to explore a visualization by requesting a summary (four finger tap) or triggering exploration (long press). Then, to request more information, they can hover their finger over a datapoint, or over a physical marker. To cancel an audio cue, they can tap on the screen with two fingers, and to repeat the previously played audio cue they can tap with three fingers anywhere}
    \Description{The data visualization prototype. A scatterplot data visualization is displayed with a cutout tactile overlay. There is an x-axis and a y-axis with tactile markers scattered across the bottom and left side of the prototype in landscape mode. The graph displays all the gestures that can be done in the prototype: Long press (start exploration), hover one finger over data (reads data point), two finger tap (stops audio), three finger tap (repeat last audio), and four finger tap (provides overview of data visualization).}
    \label{fig:dataviz}
\end{figure}

\subsubsection{Braille Raised Markers Overlay} \label{subsubsec:tactile_overlays_braille}

For this overlay, we wanted to evaluate to what extent users' prior knowledge of Braille would be helpful for navigation. While researchers have found that sighted users recognize cutout markers more accurately than raised markers\cite{kalia2014perception}, these explorations only involved blindfolded sighted people and used abstract shapes for their evaluation. Thus, we wanted to remain open to both alternatives, considering the expertise of Blind people with raised Braille. We expected users would navigate quicker by reading the Braille cells nearly instantaneously. Our Braille markers followed the Braille Authority of North America standard \cite{Brailleauthority} guidelines for creating Braille-labeled material (See Fig. \ref{fig:markerdesign}). This resulted in an overlay with 21 rows by 14 columns (used in portrait mode).

\subsubsection{Geometric Shape Cutout Markers Overlay} \label{subsubsec:tactile_overlays_cutouts}

We wanted to create tactile touch-points accessible to both Braille and non-Braille users. 
To this end, we decided to use geometric shapes, as most users would be familiar with these shapes, and low-angle polygons have contours easier to discern by touch. Cutout marker shapes included five elements in a repeating pattern: a circle, two diamonds in sequence, a triangle, a square, and a pentagon. Based on our pilot study and internal testing, we set the markers sizing to 0.75 cm by 0.75 cm (See Fig. \ref{fig:markerdesign}), and the markers were separated by 1 cm (measured from the center of each marker). This resulted in an overlay with 14 rows and 25 columns (used in landscape mode).

In addition, we included a line cutout every five elements to divide the screen into ``quadrants''. The system leveraged these quadrants to serve as physical landmarks and chunk and summarize information to users when triggering screen reader prompts. See Table \ref{tab:tactile_design} for more details.

\begin{figure}
    \centering
    \includegraphics[width=1\linewidth]{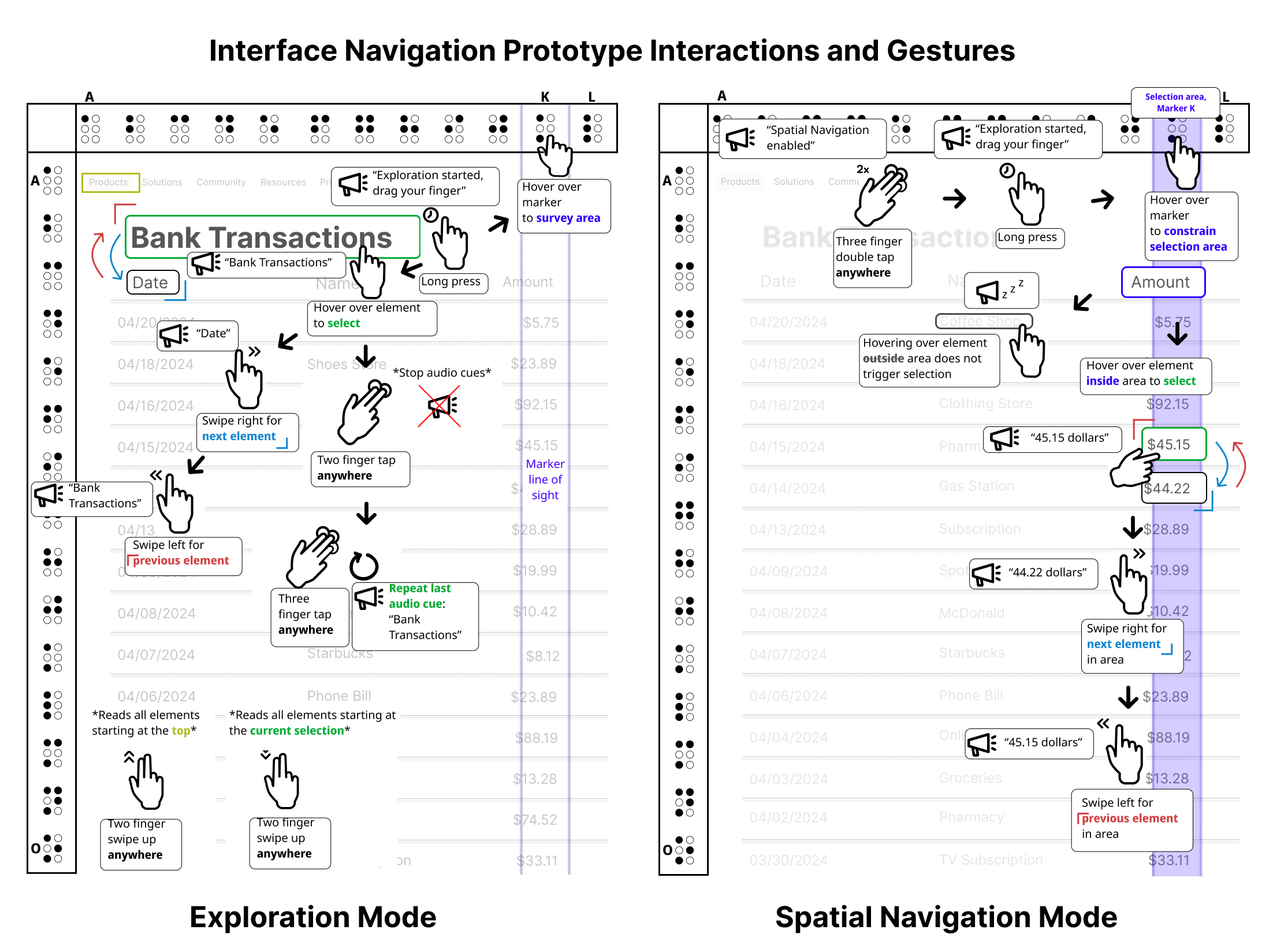}
    \caption{Both exploration mode (on the left) and spatial navigation mode (on the right) in the interface navigation prototype used similar gestures as the Data Visualization prototype but included new swiping gestures to control navigation. The spatial navigation mode also showcases how participants select specific interaction areas by touching a physical marker to constrain their navigation inputs. In this example, when the user touches the Coffee Shop label, the screen reader does not provide any auditory feedback because it is outside the currently selected screen division of the Braille letter K (Column 11). If the user wants to access this label, they would either select Row "D" or Column "F" which have line of sight of this label, or they can go on exploration mode.}
    \Description{The interface navigation prototype. A bank transactions application is displayed with a braille tactile overlay. There is an x-axis and a y-axis with tactile markers scattered across the top and left side of the prototype in portrait mode. On the left side, the graph displays all the gestures that can be done in the prototype: Hover over a marker to survey the marker line of sight, long press to trigger exploration, swipe right or left to read the next element or previous element, two finger tap (stop audio), three finger tap (repeat the last audio), two finger swipe up and down. On the right side, the graph displays all the gestures that can be done in the spatial navigation mode: Hover over marker to lock selection to elements on line of sight, hover over elements inside the area, swipe right or left to shift focus to next or previous elements on screen, and double tap with three fingers to enable or disable spatial navigation mode.}
    \label{fig:gestures_general}
\end{figure}

\subsection{Screen Reader Experience} \label{subsec:screenreadergestures}

\input{tables/gestures}

\input{tables/gestures_table}

In our study, we developed two distinct screen reader experiences tailored to specific prototypes: the data visualization prototype and the interface navigation prototype, each with unique speech prompts and interaction gestures to align with their goals (See Tables \ref{tab:normalgestures}, and \ref{tab:gestures}).

\subsubsection{Data visualization prototype gestures and feedback:}
The data visualization prototype was designed to provide users with an overview of the visualization, focusing on quantitative variables encoded on each axis. It employed simple gestures to facilitate interaction. A single tap on the canvas axes allowed users to receive on-demand information about the current scale, including minimum and maximum values, step increments, and data dimensions. Using one or two fingers, a long press activated exploration by touch, enabling users to pan their fingers around the canvas to explore markers and data points. Hovering over markers provided information about marker numbers, data points, min/max values, and quadrants, while hovering over data points triggered auditory cues. Lifting a finger provided a summary of selected data points. Additional gestures included a two-finger tap to cancel ongoing auditory cues or speech prompts, a three-finger tap to replay the last spoken prompt, and a four-finger tap to summarize the current data visualization, including total data points and axis details. See figure \ref{fig:dataviz} for more details.

\subsubsection{Interface navigation prototype gestures and feedback:}
Conversely, the interface navigation prototype emphasized exploration through familiar gestures based on standard screen reader software gestures like VoiceOver \cite{apple_voiceover} and TalkBack \cite{google_talkback}. Using one or two hands, a long press with one finger activated exploration by touch, allowing users to focus on-screen elements, with a ``tick'' sound and element label announcement when crossing elements. The two-finger and three-finger taps functioned similarly to the data visualization prototype. A one-finger swipe right advanced the virtual cursor to the next element, with auditory cues indicating focus changes and a ``thonk'' sound when reaching the end of a list. In contrast, a swipe left moved the cursor to the previous element with similar feedback. A two-finger swipe up initiated a summarized reading of all screen elements starting from the first one, and a swipe down continued reading from the current cursor position. A double tap with three fingers enabled or disabled spatial navigation mode. See figure \ref{fig:gestures_general} for more details.

\subsection{Spatial Navigation} \label{subsec:spatialnavigation}

\input{tables/spatial_nav}

Exclusive to the interface navigation prototype, spatial navigation mode introduced a way to navigate screen elements with spatial constraints. Users could select a row or column, limiting interactions to elements within that group. This mode altered the effect of certain gestures (See Table \ref{tab:spatialgestures}). For example, a long press allowed users to choose a row or column by hovering over a marker, focusing the virtual cursor on the first item in the group, with the screen reader only reading elements within the selected group. Swipes right and left navigated elements within the selected group, announcing the item's order. The two-finger swipes up and down were disabled in this mode, while the double tap with three fingers toggled spatial navigation mode on or off. This design allowed users to navigate and understand the spatial layout of screen elements efficiently, enhancing their interaction experience. See figure \ref{fig:gestures_general} for more details.

\subsection{Implementation} \label{subsec:implementation}

\begin{figure}
    \centering
    \includegraphics[width=0.74\linewidth]{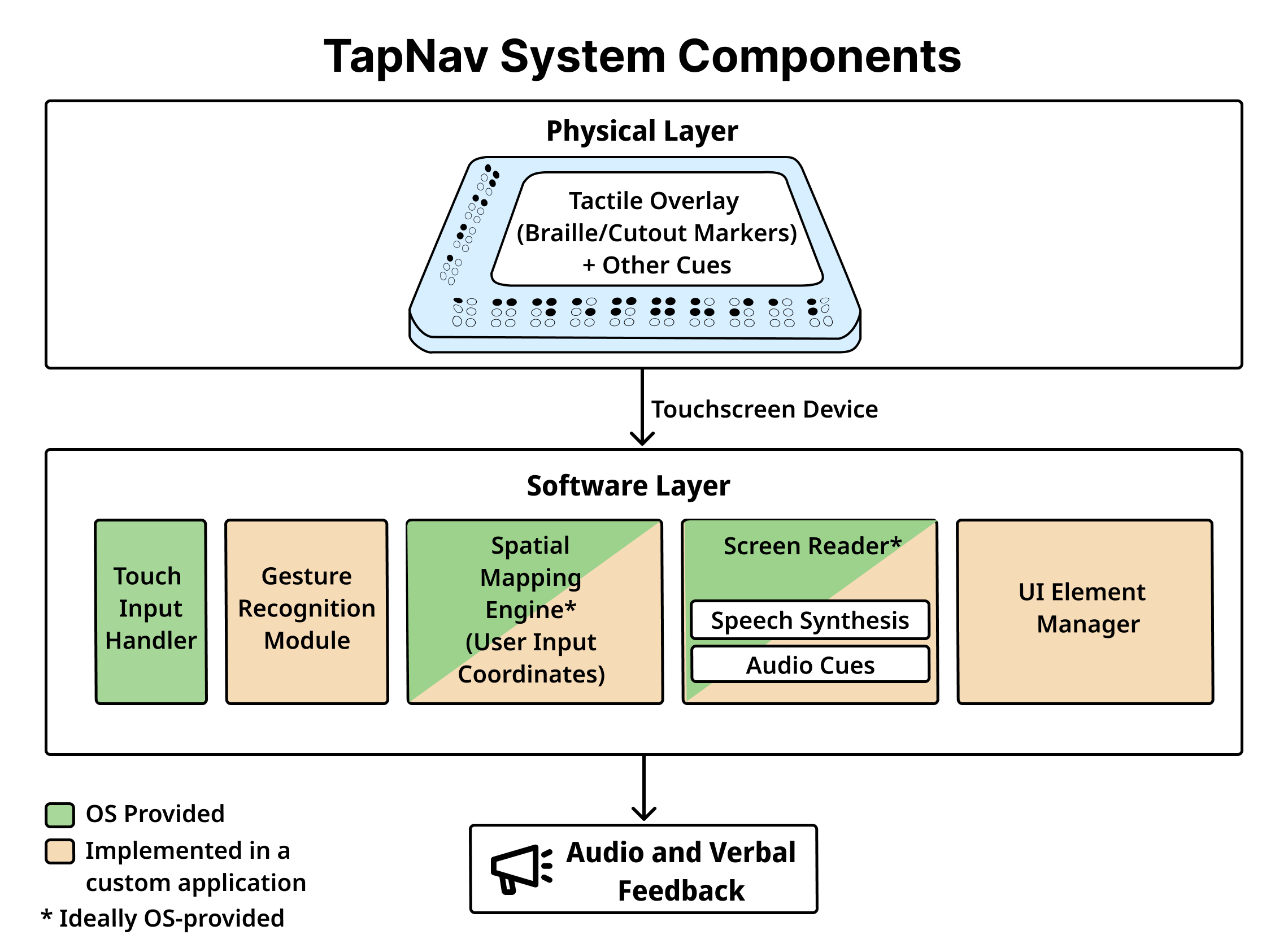}
    \caption{Components of TapNav: BLV users touch a physical overlay placed on a touchscreen device. Then, through gesture recognition, and a spatial mapping engine, the screen reader module generates adaptive auditory feedback. Green modules denote OS-provided components, while orange modules indicate components we implemented for our prototype.}
    \Description{A diagram displaying the components of the TapNav System: a physical layer (which is a tactile overlay),  a software layer (which is comprised by a touch input handler, gesture recognition module, a spatial mapping engine, a screen reader, and a UI element manager), and an output layer (adaptive audio and verbal feedback).}
    \label{fig:tapnavcomp}
\end{figure}

\DataTap{} is a standalone application developed using the open-source, cross-platform framework React Native~\cite{reactnative_github}. This choice enables deployment on both iOS and Android devices. Our final prototype targeted an Android tablet, but \DataTap{} is portable to iOS.

\textbf{System components.}
TapNav comprises a physical layer (tactile overlay on a touchscreen), and a user input processing layer (gesture recognition, and spatial mapping engine), and an output feedback layer (UI elements display, speech synthesis and audio cues via a custom screen reader). See Fig.~\ref{fig:tapnavcomp} for more details.

To create the overlays, we followed two processes: (1) the Braille overlay was produced with Braille labeling tape and a standard Braille embossing labeler; and (2) the cutout overlay was produced with vinyl transfer tape using a Silhouette Cameo 4 cutting machine.


%% file: tables/tactile.tex
\begin{table}[t]
\centering
\renewcommand{\arraystretch}{1.25}
\setlength{\tabcolsep}{6pt}
\begin{tabular}{
     >{\raggedright\arraybackslash}m{0.35\textwidth}|
    >{\centering\arraybackslash}m{0.18\textwidth}|
     >{\centering\arraybackslash}m{0.18\textwidth}|
     >{\centering\arraybackslash}m{0.18\textwidth} 
}

\textbf{Tactile Overlay Design} & \textbf{PoC (Pilot)} & \textbf{Data Visualization Prototype} & \textbf{Interface Navigation Prototype} \\
\hline
Cutout Shape Markers & $\bullet$ & $\bullet$ &  \\

Braille Raised Markers &  &  & $\bullet$ \\

Axis Grid Alignment (X/Y Reference) & $\bullet$ & $\bullet$ & $\bullet$ \\

Quadrant Markers for Major Screen Divisions & $\bullet$ & $\bullet$ &  \\

\end{tabular}
\caption{Tactile overlay design features across prototypes. A bullet ($\bullet$) indicates the feature was implemented for the overlay used in the prototype.}
\label{tab:tactile_design}
\end{table}

%% file: tables/gestures.tex
\begin{table}[h]
\centering
\begin{tabular}{>{\raggedright\arraybackslash}p{2.5cm}|>{\raggedright\arraybackslash}p{6cm}|>{\raggedright\arraybackslash}p{6cm}}

\textbf{Gestures} & \textbf{Data Visualization Prototype} & \textbf{ Interface Navigation Prototype} \\ \hline
Single Tap on Canvas Axes & Provides scale information: min/max values, step increments, and data dimensions. & N/A \\ \hline
Long Press (One/Two Fingers) & Activates exploration by touch; provides marker and data point information. & Activates exploration by touch; focuses on screen elements with auditory cues. \\ \hline
Tap with Two Fingers & Cancels ongoing auditory cues or speech prompts. & Cancels ongoing auditory cues or speech prompts. \\ \hline
Tap with Three Fingers & Replays the last spoken prompt. & Replays the last spoken prompt. \\ \hline
Tap with Four Fingers & Provides a summary of the data visualization: total data points, axis details. & N/A \\ \hline
Swipe Right with One Finger & N/A & Advances the virtual cursor to the next element; provides auditory feedback. \\ \hline
Swipe Left with One Finger & N/A & Moves the virtual cursor to the previous element; provides auditory feedback. \\ \hline
Swipe with Two Fingers Up & N/A & Initiates summarized reading from the beginning of the screen elements. \\ \hline
Swipe with Two Fingers Down & N/A & Continues reading from the current cursor position. \\ \hline
Double Tap with Three Fingers & N/A & Enables or disables spatial navigation mode. \\ 
\end{tabular}
\caption{Gestures and their functions in the Data Visualization and Interface Navigation Prototypes}
\label{tab:normalgestures}
\end{table}

%% file: tables/gestures_table.tex
\begin{table}[t]
\centering
\renewcommand{\arraystretch}{1.25}
\setlength{\tabcolsep}{6pt}
\begin{tabular}{
     >{\raggedright\arraybackslash}m{0.35\textwidth}
    | >{\centering\arraybackslash}m{0.18\textwidth}
    | >{\centering\arraybackslash}m{0.18\textwidth}
    | >{\centering\arraybackslash}m{0.18\textwidth} 
}

\textbf{Gesture} & \textbf{PoC (Pilot)} & \textbf{Data Visualization Prototype} & \textbf{Interface Navigation Prototype} \\ 
\hline
Single Tap on Axis & $\bullet$ & $\bullet$ &  \\

Single Tap on Canvas & $\bullet$ & $\bullet$ & $\bullet$ \\ 

Long Press (One or Two Fingers) & $\bullet$ & $\bullet$ & $\bullet$ \\

Two-Finger Tap (Cancel Speech) & $\bullet$ & $\bullet$ & $\bullet$ \\

Three-Finger Tap (Replay Last Prompt) & $\bullet$ & $\bullet$ & $\bullet$ \\

Four-Finger Tap (Summary or Overview) & $\bullet$ & $\bullet$ &  \\

One-Finger Swipe Right or Left (Navigation) &  &  & $\bullet$ \\

Two-Finger Swipe Up or Down (Continuous Reading) &  &  & $\bullet$ \\

Three-Finger Double Tap (Toggle Spatial Navigation) &  &  & $\bullet$ \\

\end{tabular}
\caption{Gesture availability and functionality across prototypes. 
A bullet ($\bullet$) indicates the gesture was implemented and available.}
\label{tab:gestures}
\end{table}

%% file: tables/spatial_nav.tex
\begin{table}[ht]
\centering
\begin{tabular}{>{\raggedright\arraybackslash}p{4cm}|>{\raggedright\arraybackslash}p{8cm}}

\textbf{Gestures} & \textbf{Function in Spatial Navigation Mode} \\ \hline
Long Press (One/Two Fingers) & Selects a row or column; focuses on elements within the group; provides spatial feedback. \\ \hline
Swipe Right with One Finger & Navigates elements within the selected group; announces item order. \\ \hline
Swipe Left with One Finger & Navigates elements within the selected group; announces item order. \\ \hline
Double Tap with Three Fingers & Toggles spatial navigation mode on or off. \\ 
\end{tabular}
\caption{Gestures and their functions in Spatial Navigation Mode ( Interface Navigation Prototype Only)}
\label{tab:spatialgestures}
\end{table}

%% file: src/06_evaluation.tex
\section{User Study}

To evaluate the potential of tactile overlays and spatial navigation to support touchscreen navigation across different types of visual content, we conducted a user study of \DataTap{} with BLV people testing two distinct prototypes: One for data visualizations, and one for general-purpose interface navigation. 

All methods and procedures were reviewed and cleared by our organization's relevant compliance and control functions. 

\subsection{Participants}
\label{subsec:participants_main}
\input{tables/demographics} 
We recruited participants through a third-party service that works directly with Blind and Low Vision communities. The third-party service collected informed consent from all participants. All participants were located in the USA. To be eligible for our user study, participants needed to self-identify as Blind or Low Vision, have experience using screen readers such as VoiceOver or TalkBack, be 18 years or older, and be located within physical travel distance of our location. Participants were compensated through the third-party service.

Table \ref{tab:demographics}, presents the demographics of our participants (n=12, 4 female, and 8 male). Participants age ranged from approximately 20 to 70 years old (median = 52, IQR = 22), with 8 participants identifying as blind and 4 as having low vision.

\subsection{Procedure}

\begin{figure}
    \centering
    \includegraphics[width=1\linewidth]{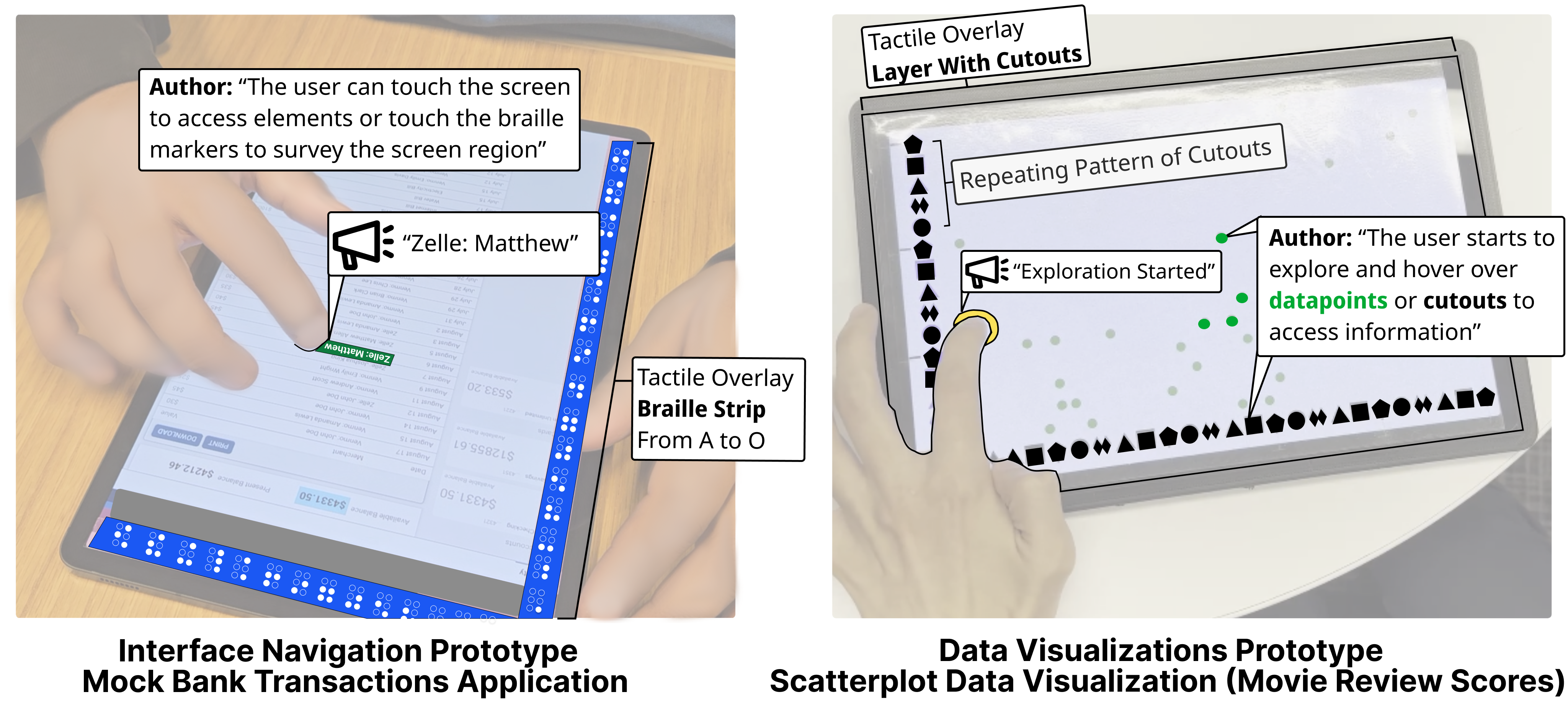}
    \caption{Participants experienced two prototypes to examine the use of \DataTap{}, our \textit{adaptive spatiotactile screen reader}: a mock bank transactions application (left) and a scatterplot data visualization (right). A different tactile overlay is placed on a touchscreen device in both scenarios: a Braille strip and a layer with cutouts. Each physical marker in the tactile overlay corresponds to specific digital content to reveal underlying structure to BLV people.}
    \Description{Participants experienced two prototypes to examine the use of TapNav, our \textit{adaptive spatiotactile screen reader}: a mock bank transactions application (left) and a scatterplot data visualization (right). A different tactile overlay is placed on a touchscreen device in both scenarios: a Braille strip and a layer with cutouts. Each physical marker in the tactile overlay corresponds to specific digital content to reveal underlying structure to BLV people.}
    \label{fig:proto-images}
\end{figure}

We conducted a one-hour in-person session to test \DataTap{}, recording audio and participant screens. Participants interacted with two prototypes: a scatterplot data visualization prototype using a cutout-based tactile overlay and a interface navigation prototype using a Braille-based tactile overlay for a mock bank transactions application. In a within-subjects study, each participant completed tasks using both prototypes (See Fig. \ref{fig:proto-images}). To mitigate learning effects, we balanced the order of prototype exposure and used two different datasets to address potential biases in data arrangement and transaction location.

The in-person session was divided as follows:
\begin{itemize}

    \item The prototype introduction covered novel gesture inputs, how to ``read'' the tactile overlay, and included free exploration with an unrelated tutorial scenario for familiarization. This lasted between 10 and 15 minutes. For the data visualization scenario, participants experienced a test dataset. For the  interface navigation prototype, participants experienced a pdf document with paragraphs, tables, and lists.

    \item Participants completed interaction tasks by searching for information on the screen and reporting locations using row and column coordinates. For the data visualization prototype, tasks included a look-up (e.g., find ``Avengers'' in column 24, row 13), a localization (e.g., see ``The Circus''), and a browse task (e.g., find the highest-rated film). For the interface navigation prototype, tasks included a browse (e.g., find a transaction over \$50) and an exploration task (e.g., identify a fraudulent transaction if you only own cars). These tasks took approximately 10 minutes.

    \item A short semi-structured questionnaire where we asked participants about the prototype features, such as the usefulness of the current tactile overlay (e.g., Braille vs Cutouts), the screen reader experience, and the spatial navigation mode. This lasted between 5 to 10 minutes.

    \item These steps were then repeated for the second prototype, often requiring less time for the introduction and familiarization. 

    \item A closing semi-structured interview where we asked participants' overall thoughts about the tactile overlays (e.g., how they compared to one another), their potential use for other scenarios, and feedback on how to improve future multimodal touchscreen systems that apply tactile overlays. This lasted approximately 5 to 10 minutes.
\end{itemize}

\subsection{Analysis}

We followed an inductive open coding process to analyze the study sessions and interviews, then used two rounds of affinity diagramming to group codes into themes \cite{hartson2012ux}. 


One researcher coded three sessions to generate an initial codebook, which was then revised by three researchers and condensed to focus on aspects related to spatial navigation, screen reader and gestures design, and tactile overlays. 
Finally, three researchers used the revised codebook to code one additional session, compared their codes, and validated the revised codebook. 
One researcher coded the remaining eight sessions.
After finishing the coding process, we conducted two rounds of affinity diagramming, organizing coded interview snippets by their thematic similarity into groupings to develop formal themes.
One researcher performed the initial process, and then three researchers revised the proposed themes and reorganized the coded snippets as needed during a second affinity diagramming session. The resulting themes are presented in the findings.

%% file: tables/demographics.tex
\begin{table}[!ht]
\centering
\caption{Participant Demographics}
\label{tab:demographics}
\begin{tabular}{@{}llllll@{}}
\toprule
\textbf{PID} & \textbf{Vision} & \textbf{Braille Experience} & \textbf{Gender} & \textbf{Age} & \textbf{Assistive Tech Used} \\
\midrule
P1  & Low Vision & No experience          & M & 65+    & JAWS, VoiceOver \\
P2  & Blind      & Experienced            & F & 20--39 & JAWS, VoiceOver, ZoomText \\
P3  & Blind      & Experienced            & M & 20--39 & NVDA, VoiceOver \\
P4  & Blind      & Experienced            & M & 40--64 & JAWS, VoiceOver \\
P5  & Blind      & Advanced experience    & F & 40--64 & JAWS, VoiceOver, NVDA \\
P6  & Blind      & Experienced            & M & 20--39 & JAWS, VoiceOver \\
P7  & Blind      & Advanced experience    & M & 20--39 & JAWS, VoiceOver \\
P8  & Low Vision & Basic experience       & M& 40--64 & JAWS, Window-Eyes \\
P9  & Blind      & Experienced            & M & 65+    & JAWS, VoiceOver \\
P10 & Low Vision & Advanced experience    & F & 20--39 & JAWS, NVDA, VoiceOver, Talkback \\
P11 & Blind      & Experienced            & F & 40--64 & JAWS, VoiceOver \\
P12 & Low Vision & No experience          & M & 40--64 & JAWS, VoiceOver \\
\bottomrule
\end{tabular}
\end{table}

%% file: src/07_findings.tex
\section{Findings}

We derived three major themes from our thematic analysis touching on key aspects of \textit{adaptive spatiotactile screen readers}: tactile overlays, spatial navigation, and screen reader experience.

\subsection{Tactile Overlays}

In this section, we present participants' insights to understand how to help them better localize content and perform movements with the guidance of tactile cues, and additional considerations to inform the design of tactile markers (e.g., Low Vision versus Blind users needs, perception challenges, etc.).

\subsubsection{Supporting Localization, and Alignment}

Participants discussed how the tactile overlays supported navigating, localizing reference points with the tactile makers, and how tactile cues could support user alignment.
Eight participants found the tactile overlays helpful to recognize specific screen regions (P1, P2, P3, P4, P5, P7, P9, and P12). P9 explained that for the prototype with the Braille overlay, it is easy to recognize a column or row because, \textit{``if you know Braille, you know automatically, OK, this is column 10 because J is the 10th letter.''} P7 shared a similar sentiment for both tactile overlays: \textit{``The overlays were very helpful for my navigation just because they give me a sense of which row and column I'm in.''} P7 further contrasted their experience with their standard interactions on touchscreen-based devices, noting \textit{``some of the other uses that I've experienced before where it's just a smooth surface (...) it feels sort of arbitrary.''} Users differentiated specific screen locations by having distinct tactile touch points to support them during navigation tasks.

While participants found the tactile overlays helpful in locating specific screen regions, we observed that all participants except P7 found challenging to do vertical and horizontal line movements toward the middle of the screen. This primarily affected participants' interactions with the data visualization prototype. P12 subsequently preferred the interface navigation prototype: \textit{``When I flicked, it gave me navigation. Whereas with the [data visualization prototype] If I didn't go absolutely straight (...) I may or may not find the data points that I needed.''} In the interface navigation prototype, participants used spatial navigation mode, touched the tactile markers, and then swiped right or left to survey through the elements in that screen region if they were struggling with alignment (or used the filtered exploration by touch). However, for the data visualization prototype, participants only option was to use exploration by touch and move towards the middle of the screen to localize the data points with no constraints. Three participants recommended adding horizontal and vertical raised guidelines aligned with markers to support vertical and horizontal movements (P7, P9, and P10). P9 explained that the lines could \textit{``give them a spatial idea of how it is set up''}, and that he can just \textit{``glide and slide, and glide (...) to do things much faster.''} While tactile guidelines could be explored to support user movement alignment, P8 also warned that too many tactile cues could render tactile sensing confusing and overwhelming.

\subsubsection{Blind Versus Low Vision People Preference for Tactile Markers} \label{sec:considerations}

Participants discussed their preferences between the Braille and cutout markers and how each design might be preferable based the person's knowledge and health condition. Most Blind participants preferred Braille and raised lines over the cutouts (P2, P3, P6, P10, and P11) because they were accustomed to Braille. P3, after trying the data visualization prototype with cutout markers first, said that having \textit{``raised markers''} would be \textit{``a lot easier, especially for visually impaired users''} to perceive the markers. P11 supported this sentiment, saying that the Braille is \textit{``much easier [to read] because it is a formal and more visible marker.''} They further noted that the cutouts were \textit{``harder to feel through and distinguish. You literally had to count through [them]''}.

While some Blind participants found the cutouts dissatisfying, P1 and P12, who were Low Vision and non-Braille users, found both the cutouts and the Braille markers useful. What P11 found negative about the cutouts (e.g., having to count through markers at times), P12 highlighted as a positive for the Braille overlay they could not read: \textit{``even though I can't read Braille, I know that if I went to a particular row counting down and then count the letters (...) I will know where I am on the screen.''} P1 also appreciated how the Braille overlay could be accessible to him and his Blind allies: \textit{``What you've done is important for a person who reads Braille, they can navigate (...) and they'll know exactly where they are. For someone who does not know Braille like myself, it still gives you feedback.''} 

Two participants preferred cutout markers over Braille markers (P8, P12). P8, who is a basic user of Braille, explained that the cutout markers are \textit{``better for someone that has less dexterity or sensitivity,''} and added, \textit{``if they don't know Braille, [the Braille overlay] might be a little intimidating because not all blind people know Braille.''} P8 also added that raised Braille cells can become worn out over time, making it difficult to perceive such markers. On the other hand, P12, who cannot read Braille, preferred the cutouts because for the Braille markers, they \textit{``can't tell the difference between an A and an N.''} However, they reiterated that having \textit{``touch points''} (i.e., any tactile marker) does have functionality even if they cannot read them.

\subsubsection{Tactile Perception Challenges and Design Recommendations for Tactile Markers}

Participants discussed challenges in perceiving tactile markers, which cutout markers were easiest to perceive, and proposed recommendations for designing future tactile markers. Eight participants felt that some of the cutout markers were difficult to discern from one another (P2, P5, P6, P7, P8, P10, P11, and P12). For instance, three participants explained that the pentagon and circle markers felt too similar (P7, P8, and P10). P8 explained, \textit{``I really can't tell that's a pentagon at all. (...) I don't feel the angle of it being a pentagon.''} P10 speculated, \textit{``Maybe it's just a little too finite for me or something''} trying to explain why it was difficult to perceive the pentagon. To improve the legibility between these two shapes P8, and P10 recommended to have figures sides \textit{``sharper and more defined''} or having a \textit{``thicker overlay''}. Another shape that was difficult to perceive was the double diamond. Three participants mentioned they did not feel the marker as a double diamond (P5, P6, and P8). P5 said they would not have known the shape was a double diamond if they were not told during the tutorial session.

During the user study, we observed the majority of participants were most comfortable with the circle cutouts. In particular, P5, P8 and P11 said that the circle shape was very \textit{``clear''}, that they  felt it \textit{``really well''} and that it felt \textit{``good''}. As for the triangle and the square cutout markers, P5 and P8 also said they were distinguishable. In short, simpler shapes with a lower number of sides and easy to trace outlines were preferred by participants. P8 recommended other potential simple shapes to replace the double diamond and the pentagon, such as a semi circle in the shape of a ``D'', or a half moon shape. P8 added that some of these shapes could be rotated if more distinct shapes were needed, such as an upside-down triangle but recommended avoiding having two similar shapes in sequence (e.g., a triangle and then an upside-down triangle).

For the Braille markers, participants discussed the Braille cells' orientation and their expectations for a grid-based system. P3 explained that reading Braille vertically is difficult because they \textit{``appear different''} when dragging their finger from top to bottom: \textit{``Like, you have to kind of switch your thinking (...) is it an M or an U? It's a U but it's sideways.''} While only P3 explicitly mentioned the added cognitive load of reading Braille vertically (e.g., turned 90 degrees clockwise), other participants like P8 turned around the device sometimes to read the Braille markers horizontally.

Two participants also highlighted how health conditions can affect  tactile perception skills (P4, P8). Specifically, P4 warned that many BLV people experience vision loss due to diabetes and expanded that people with diabetes are typically affected by neuropathy too, a condition that affects their fingertip sensitivity. While P8 agreed that users with lower fingertip sensitivity may struggle to perceive tactile markers, they believed that the cutout markers might be easier to discern than the Braille cells because \textit{``it might be hard to feel the little dots, whereas those cut outs are easier to feel.''} 

Some participants provided specific recommendations to leverage users' prior knowledge using Braille labeled materials with other grid-based systems, such as accessible math tactile graphics, Excel, and popular board games like Battleship (P3, P5, P7, P9, and P10).  In particular, participants explained that one axis should be labeled as letters and the other as numbers to find specific coordinates easier. P5 also added the ordering of Braille cells was unexpected: \textit{``when you said the top is row 15, I found it confusing. (...) I am going based off excel spreadsheets layout, the numbers of rows start from the top.''} While most participants did not raise any issues with the use of letters for both sides and the location of where the index starts the count (e.g., top vs bottom), it is important to consider preexisting standards for grid-based navigation and orientation to minimize users' cognitive load.

Finally, some participants proposed simplifying the markers altogether to optimize for quick counting rather than for legibility of individual markers (P9, P11, and P12). P12 explained when they did not necessarily recognize some of the markers, they could \textit{``count it out''} and \textit{``felt the tactile [markers] going through.''} P11 wanted to have \textit{``distinguished''} and repetitive markers if they are going to have to count, something like short strikes or \textit{``little boxes''} to make it easier to count across \textit{``up and down.''} While P9, and P12 did not recommend strikes or \textit{``little boxes''}, they mentioned having something simple like round bumps would make it so that they \textit{``do not worry about what shape it is.''} P9 specified: \textit{``It would be easier to use to a lot of people, and much faster, much faster navigation.''}

\subsection{Spatial Navigation}

In this section we present participants perspectives of how the spatial navigation mode enabled systematic, faster and precise exploration of screen content for certain tasks during the user study. 

\subsubsection{Spatial Navigation Mode Affords Speed, and Precision}

Seven participants found the spatial navigation mode helpful to access content faster (P2, P4, P5, P6, P8, P9, and P11). The Braille markers generally allowed users to quickly \textit{``read''} and access specific elements by row or column. For example, P6 explained that the Braille markers worked as \textit{``shortcuts''} to access information faster. P8 added that the spatial navigation mode allowed them to feel \textit{``more in charge, more in control (...) like I could go fast, and hear any information.''} Relatedly, four participants found the spatial navigation mode and the overlay allowed them to be precise with their inputs, in contrast to traditional screen reader navigation without tactile cues (P1, P5, P7, and P11). P2 and P7 said that the tactile overlay was \textit{``pretty helpful''} because \textit{``they knew exactly where each thing was supposed to be [on the screen].''} P5 added, \textit{``it is easy to move from one element to another dragging the finger, and if you press the Braille, it is precise. So it would shorten the [navigation] time.''} In short, participants used the markers as anchor points to orient and locate themselves faster. At the same time, the auditory feedback confirmed the location of their finger with respect to the content.

While participants perceived that the spatial navigation mode enabled faster and more accurate user inputs in the interface navigation prototype, six participants found it difficult to triangulate data points in the data visualization prototype due to alignment challenges (P7, P8, P9, P10, P11, and P12). P7, who had extensive experience with tactile graphics, used his personal experience to deal with this challenge: \textit{``I always have to use other fingers, and keep my hand at the same shape at all times as if playing the piano, but (...) not in one dimension but in two dimensions.''} P7 used a unique two-hand position to perform straight-line movements, showing potential alternatives to increase precision while staying aligned. Despite this, P7 clarified that \textit{``it's still kind of difficult,''} and later added that including a tactile cue like a strike (e.g., raised lines) would increase his speed and accuracy. In short, both prototypes conveyed how screen content was laid out spatially. However, without the spatial navigation mode in the data visualization prototype, participants were missing support to stay aligned with the markers (e.g., by feedback or system functions). Such support would allow users to access data points and screen elements in the middle of the screen as quickly as when reading the tactile overlay.

\subsubsection{Spatial Navigation Affords Structure}

Seven participants found that the combination of the spatial interactions and tactile markers allowed them to access content in an \textit{``organized''}, \textit{``filtered''} or \textit{``systematic''} way (P3, P5, P7, P8, P9, P11, and P12). P3 explained the spatial navigation mode helped him skim through the content: \textit{``You immediately know what you are gonna be getting once you're in a column.''} P3 also said that spatial navigation helped when there are too many elements on-screen: \textit{``The issue of trying not to like get overwhelmed by how tightly packed everything is, is less of an issue.''} P7 expressed similar sentiment, appreciating being able to navigate the data visualizations by screen regions (e.g., rows and columns): \textit{``I liked being able to navigate the data sets and be able to analyze certain groups of points (...) and figure out where things were located.''} P11, while touching different Braille markers, explained how they used the rows and columns to sort and filter what they are looking for: \textit{``These are all my transactions and this is where I find my balance, (...) I could do it by date or I could do it by amount.''} Thus, the spatial navigation mode allowed users to create a mental map of the interface, consume grouped information, and guide their navigation patterns.

Both Blind and Low Vision users were able to create these mental maps, and scaffold efficient exploration. P11 and P12, while having different vision impairments and experience with Braille, both thought the spatial navigation mode allowed them to \textit{``work their way through''} and helped them \textit{``search or understand what is on the screen more systematically.''}  Similarly, P5, and P11 leveraged the spatial navigation mode to navigate efficiently using their spatial knowledge of the interface content. P5 said: \textit{``Not just quickly but accurately find the items, instead of like dragging around the whole screen. Even if I know where college savings is with just ordinary navigation, I have to like aim towards there. Kind of feel around, and maybe takes me five gestures. With the Braille [overlay and spatial navigation] it is one tap.''} 

Finally, P5 shared unique thoughts about how spatial navigation could support collaboration and communication with mixed abilities teams, such as sighted colleagues. P5 said: \textit{``I think beyond [accuracy and efficiency], it would give you a sense that I feel equal to everyone else.''} When asked why, she explained that she can prompt her colleagues with specific information like coordinates, making her \textit{``more efficient''} and \textit{``more intelligent''}: \textit{``Because if I say to someone look at flower 30. And I tell them it's column 25 out of 13 that they can find it there. (...) So I have equal amount of information access to information that a person who sees.''} Similarly, spatial navigation would allow P5 to take a more proactive approach to look up data points or screen elements when prompted by a colleague: \textit{``if they're looking at row 24 in column 13,  this tactile overlay lets me know that (...) I'm at the same place as they are.''} Thus, the tactile overlay and the spatial navigation can bridge understanding between screen reader and sighted users. While Blind and sighted users' experience of accessing interfaces might be significantly different, spatial and tactile reference points could bridge the gap so both parties understand what the other can \textit{``see.''}

\subsection{Screen Reader Experience}

In this section we summarize participants perspectives regarding the design of the gestures we implemented for TapNav, the importance of the exploration by touch gesture to enable reliable real-time screen content exploration, and participants suggestions for alternative navigable structures to organize screen reader content.

\subsubsection{Gestures Felt Familiar and Swiping is a Core Gesture for Navigation} \label{find:gestures}

Six participants found some of the gestures familiar, and easy to use (P1, P3, P4, P8, P10, and P12). Participants pointed out the swiping gestures implemented for the interface navigation prototype were easy to perform and intuitive. For example, P10 said: \textit{``It is pretty much in keeping with gestures that most people already know.''} Interestingly, during the gestures tutorial, some participants performed gestures without receiving instructions from the researchers. For example, P10 performed the gesture to stop the screen reader audio while learning the explore by touch gesture.  When asked how they knew the gesture and its function, they responded: \textit{``I know that's how I do it with the iPhone. So I just thought what the hell I'll try it (...) So that is very intuitive.''}P1, P10 and P12 shared similar thoughts and found that other tapping-based gestures like to silence the screen reader and toggle spatial navigation mode were \textit{``simple''} and \textit{``pretty close to standard VoiceOver gestures.''}

During testing of the data visualization prototype, six participants desired the ability to scroll through data points by swiping (P2, P3, P6, P9, P10, and, P12). P12, who experienced the interface navigation prototype first, explained they missed having the ability to scroll through screen elements when selecting a screen region column or row: \textit{``Because on the first one you gave me, you could scroll. [In the data visualization prototype], you didn't quite have the same ability to scroll through to find something in a quicker way.''} Then, P3 explained the importance of swiping when using touch screens: \textit{``most of us use swiping gestures (...) [we] tend to use swiping as our primary way of navigating touch interfaces.''} In short, the swiping gesture should always be included into the interaction flow of screen reader experiences as it is one of the most basic forms of screen reader navigation users rely on.

\subsubsection{Building an Exploration by Touch Gesture}

Out of all interactions, participants engaged with the exploration by touch gesture the most since it was one of the primary ways of accessing screen content. On the data visualization prototype, it supported understanding the arrangement of data by providing summaries when touching the tactile markers, and it allowed participants to \textit{``find''} data points when touching the data visualization. On the interface navigation prototype, it enabled access to navigable structures by their location when touching the tactile markers (e.g., rows and columns), which users could further explore by touch or swiping. Consequently, participants provided extensive comments about this gesture, centering on the design of the real-time feedback generated through sound cues and verbal information when engaging in exploration by touch.

Four participants liked the \textit{``tick''} sound cue that indicated the virtual cursor changed focus while users hovered through different screen elements (P3, P4, P8, and P10). P8 explained that \textit{``it is kind of like a pause in a sentence, instead of just [reading the labels] real fast.''} Although some participants enjoyed this feedback, others felt the gesture did not work consistently (P2, P3, P6, P11, and P12). For instance, P6 said: \textit{``it felt like it wasn't reacting, (...) like it wasn't working out for me.''} For P2, the delay to trigger the gesture also made it time-consuming to use: \textit{``You have to like keep activating the explore mode. That will be really annoying when I have to like read a long email. (...) It would be pretty challenging, very time consuming.''} Additionally, P2, and P10 were confused when they hovered over empty space since the application would stay silent. P2 said: \textit{``There were a lot of spaces [with] no data. And when I wanted to find a movie, it took me a while because the screen was dead,''} referring to the lack of feedback.

Eight participants proposed recommendations to improve exploration by touch (P1, P2, P3, P5, P7, P8, P10, and P11). Participants wanted more feedback while hovering around the screen, like cues to understand when the users' finger crosses different rows or columns (P7, and P8), support mechanisms to help find and target screen elements effectively (P1, P3, P5, P10, and P11), and location information when finishing the gesture (P2 and P7). P7 and P8 proposed adding a new audio indication to communicate when users are crossing to a different region (e.g., a row or a column in our prototypes). P8 explained that this would help address alignment issues without relying solely on tactile cues: \textit{``If there was a different sound, like a musical scale, [lists out scale], if you are going the wrong way (...) you would know you are not going in a straight line.''}

P3, P5, P10, and P11 were especially concerned with overshooting targets and orienting themselves in empty space. P5 felt frustrated about trying to get more details of their target after hovering over it: \textit{``I feel like I am pretty quick [to react] but I am still overshooting. You have to be really precise.''} P1, P5 and P11 proposed adding an audio indication to help lift their finger on time (like by predicting their trajectory), or to change how the gesture works (like hovering in-place rather than having to lift the finger). On the other hand, P3 and P10 wanted to receive more feedback while navigating empty space. They proposed adding an \textit{``indicator tone''} whenever the user was moving towards a data point of interest, so that the screen does not \textit{``stay completely silent.''}

P2, and P7 proposed receiving positioning information when lifting their finger off the screen. P7 said: \textit{``that could tell you which row you're in after you hear the element.''} When asked whether they would like to receive coordinates (e.g., row and column) when hovering over an element , P7 said \textit{``Not necessarily''} and clarified \textit{``Maybe it can enabled when you're browsing, you can turn it on and off.''}

To summarize, participants found exploration by touch essential for navigation, but they highlighted key challenges related to real-time feedback, overshooting, and lack of spatial orientation cues. Their recommendations emphasize the importance of adaptive feedback, precision-enhancing mechanisms, and orientation auditory cues to ensure smooth, efficient, and confident interaction.

\subsubsection{Content, and Navigable Structures}

Seven participants found the row and column summaries to be useful for understanding how the content was laid out (P1, P2, P3, P5, P7, P10, and P12). Participants said that the content was accurate (P2), concise (P5, P7), and a great starting point when navigating (P3). P10 explained \textit{`` I like it. I think (...) once you know what you are actually listening to, then I think that it is great. It gives you the information that you are presumably after.''} For P12, when using the interface navigation prototype, touching the marker gave him initial physical confirmation he was at the right location, the content gave him a rough idea of the potential targets, and then swiping afforded fast navigation: \textit{``The summaries, you could tell exactly when, what, and how much easily [and] just by flicking around you would get what you needed.''}

Three participants proposed introducing alternative screen reader structures to complement the navigation gestures and spatial navigation for sensemaking (P1, P3, and P6). P3, and P6 proposed adding a \textit{``search bar.''} They explained that if they have an idea of what they are looking for, this feature could help users do a focused exploration and receive positional information about a datapoint or screen element they are interested about. P6 said: \textit{``So like, I could have typed in (a transaction), you know, I could have seen the transactions easier.''} Similarly, P1 explained it would be great to be able to group information by \textit{``topic''} and just \textit{``swipe''} or touch to navigate through elements specifically related to the topic. P1 gave an example with transactions related to Apple products: \textit{``For example, I wanna know everything I made with Apple. As long as I'm swiping, I could hear all the history. (...) I wouldn't have to do anything else but be sure I could stay right on Apple [related elements].''}

While row and column summaries provided a clear overview of content, participants highlighted the need for complementary navigation tools like search functions and topic-based grouping to enhance both exploration and targeted access. These findings underscore the importance of layered navigation strategies, which we further explore in section \ref{sec:discussion}.

%% file: src/08_discussion.tex
\section{Discussion} \label{sec:discussion}
Our study findings show the potential of \textit{adaptive spatiotactile screen readers} to enhance BLV users' access to touchscreen devices.   In this section, we focus our discussion on three key aspects for applying and further exploring \textit{adaptive spatiotactile screen readers}: the design of tactile overlays, the use cases of \textit{adaptive spatiotactile screen readers} for enhancing interactions on touchscreen devices, and limitations of our approach.

\subsection{Designing Tactile Overlays for Adaptive Spatiotactile Screen Readers}
    
Participants highlighted different ways tactile overlays enhanced their screen reader experience. We found that tactile markers allowed participants  locate specific screen locations using their sense of touch rather than relying solely on screen readers' feedback. Participants also provided feedback about the benefits and drawbacks of Braille and cutout markers. 

In this section, we provide a set of questions that can serve as a practical guideline for future designers to implement their own tactile overlays for \textit{adaptive spatiotactile screen readers}.

\begin{itemize}
    \item \textbf{What are the Braille Literacy levels of your users?}
    As observed from section \ref{sec:considerations}, most participants who were Blind knew Braille. This meant that they strongly preferred these types of markers for their recognizability. However, for Low Vision users or Blind users with a basic understanding of Braille (P8), the cutouts were typically perceived as better and ``less intimidating'' to use. Braille markers may also deter users who strongly identify as Low Vision as they may perceive the technology as something designed ``not for them''. Important to note, while reports of Braille literacy among BLV people are inconsistent in the mainstream\cite{sheffield2022many}, best estimates in the last two decades point to an approximate number of 19\% of Braille literacy among BLV people \cite{craig2002statewide}. Thus, Braille marker overlays may also be inaccessible for many BLV people. A solution to this problem, as pointed out by participants, is to create simpler touch-points (e.g., raised bumps) that both Blind and Low Vision users can use. Thus, depending on the target user group, it may be preferable to use Braille for Blind users, cutouts for Low Vision users or both groups, or simple touch points for both Blind and Low Vision users.

    \item \textbf{What is the use case and content dynamics?}
    We designed two tactile overlays, each tailored for a use case: data visualization exploration and general interface navigation. Each overlay incorporated distinct design choices to optimize usability within its respective context. The data visualization overlay featured a repeating pattern of tactile markers with notches to delineate quadrants, enabling experienced users familiar with mathematical tactile graphics to navigate and interpret data points more efficiently. The interface navigation overlay was designed for broader usability. It consisted of two orthogonal axes that divided the screen into rows and columns without additional segmentation. While it could function for data visualizations, it lacked features like notches, which facilitate structured faster navigation for more advanced users. As such, both overlays were flexible and constrained in their own ways. The first overlay could be enhanced with raised guidelines to improve alignment support, and while tested primarily with scatterplots, it could extend to other data visualization types. The second overlay prioritized adaptability, ensuring compatibility with diverse applications where extra tactile elements like guidelines and notches might become intrusive or distracting. Thus, the use case and the dynamics of the digital content heavily influence the design of the overlay. It may be preferable to create overlays with less tactile features (general-purpose) or more tactile features (domain-specific) depending on your goal and how those tactile features disrupt or enhance the underlying experience across different scenarios within your chosen use case.

    \item \textbf{What interactions require or would considerably benefit from tactile feedback?}
    In our interface navigation prototype, participants appreciated the ability to touch markers and swipe to navigate, as this interaction allowed them to access targets without requiring precise physical movements. Combining tactile markers and navigation gestures provided a flexible and efficient interface exploration method. In contrast, the data visualization prototype lacked navigation gestures, requiring users to make precise physical movements to access data points, particularly those positioned in the middle of the screen. This limitation led participants to recommend modifying the tactile overlay by incorporating physical guidelines to support straight horizontal movements, improving alignment and ease of access. These findings highlight how interactions can scaffold onto tactile overlays to create novel ways of guiding BLV users and providing access to information (see Section \ref{disc:usecases}). In short, tactile overlays should not be designed in isolation; instead, they should be approached through a hybrid lens, where the physical (static) structure complements the digital (dynamic) functionality based on your interaction model.

\end{itemize}

As designers answer these questions, the design of their tactile overlay should become more readily apparent. To summarize, tactile overlays improve screen reader interactions by enhancing BLV users navigation and access to screen content through touch. The effectiveness and design of tactile overlays depend on the Braille literacy of the intended target users, use case and content dynamics, and screen reader interaction model. Whether using Braille, cutouts, or simple touch points, designers must balance tactile detail with usability to avoid unintended distraction. The guidelines in this section provide an initial practical framework for designing overlays that align with user needs and integrate well with \textit{adaptive spatiotactile screen readers} experiences. 

Our findings open a design space for investigating how tactile marker perception shapes user experience with touchscreen devices. We encourage future researchers to systematically probe how marker design, texture, spacing, and material influence recognition speed, accuracy, and comfort across diverse BLV groups to further support and enhance touchscreen accessibility.

\subsection{Use Cases of Adaptive Spatiotactile Screen Readers for Interactions on Touchscreen Devices} \label{disc:usecases}

Participants felt the combination of tactile overlays and spatially constrained interactions improved their accuracy and speed for screen content navigation, highlighting how markers and tactile divisions functioned as anchoring points for exploration by touch. This enabled faster access to rows and columns while reducing reliance on sequential scanning and helped them make fewer mistakes. Prior research on spatially constrained interactions has shown similar benefits \cite{kane2011access,thompson2023chart,zhang2024charta11y}. Still, tactile overlays offer a distinct advantage: They provide physical feedback that helps users anticipate interaction outcomes while simultaneously having the option to selectively listen to auditory feedback or altogether skip it, reducing cognitive overload.

For data visualization analysis, exploration by touch was a core interaction. In this use case, participants identified several challenges: overshooting targets, lacking orientation cues in empty spaces, and receiving inconsistent feedback. To address these issues, participants recommended introducing additional auditory cues to support tasks that require alignment (e.g., tones that indicate crossing rows/columns), predictive feedback to prevent overshooting, and positional updates when lifting a finger off the screen. These recommendations suggest that data visualization analysis requires an active and dynamic navigation strategy and that it benefits from both tactile structure and adaptive feedback systems.

Beyond interface navigation and data visualization analysis, \textit{adaptive spatiotactile screen readers} can also support more specialized use cases like enhancing data storytelling for BLV users. Approaches like \citet{siu2022supporting} audio data narratives provide structured, guided experiences for understanding data visualizations. Layering tactile overlays onto audio-guided storytelling could create more interactive and useful experiences. For example, expanding on the ideas explored by the InfoSonic system \cite{holloway2022infosonics}, an adaptive audio narrative could refer to physical landmarks on the overlay to guide users through key insights, dynamically triggering screen functions or emphasizing significant trends based on touch input. This combination of audio, touch, and spatial cues could make data storytelling more immersive and structured for BLV users.

We encourage future work to imagine and explore how these systems can be extended to other applications (e.g., gaming, creative tools, or collaborative workspaces) to further broaden their impact on touchscreen accessibility.

\subsection{How Can We Extend Adaptive Spatiotactile Screen Readers?}

Although our \textit{adaptive spatiotactile screen readers} probe, TapNav, focused on a scatter-plot and a bank transactions application, the underlying mechanism—dynamically binding logical interface regions to a grid of tactile landmarks— could extend to most conventional mobile layouts (lists, cards, navigation bars, buttons, etc.) and to data views beyond 2-D plots (e.g., bar/line charts, heat-maps, etc.). Because each marker is a parametric “slot” rather than a fixed pixel, users can first indicate where their tactile markers sit (e.g., as if provided by a system-level screen reader feature), and the screen reader then dynamically re-binds interface regions, actions and speech feedback to those physical landmarks whenever the layout or orientation changes. 

Beyond user-driven manual remapping, AI could automate these bindings or scaffold additional interactions. Recent research in AI-driven data visualization analysis \cite{sharif2023understanding} enables users to query for insights such as trends, calculations, and relationships between data points. We propose integrating AI with spatially constrained interactions and markers rather than relying solely on direct query-based interaction. Generative AI interfaces can dynamically adapt to the user's context while anchoring navigation to physical landmarks on the overlay to enhance access. Instead of requiring users to query an AI with questions, the AI could generate custom navigable structures, such as grouping related data points or adjusting functional zones to match the user's tactile overlay design. This approach leverages AI's adaptability while preserving direct spatial interaction grounded in tactile feedback, ensuring that users can survey and explore data in a way that remains structured, efficient, and aligned with their preferred interaction methods.

Like in data exploration, AI-driven enhancements and tactile overlays can also complement general screen reader navigation, adapting to user needs across different interaction paradigms. Existing screen readers—whether on touchscreens or desktop interfaces—offer structured navigation mechanisms such as heading-based movement, region-based exploration, or sequential scanning \cite{kane2011access,grussenmeyer2017accessible,brown2003design}. AI can extend these strategies by providing context-aware summaries, dynamically reorganizing content, or refining navigation granularity based on user preferences. When combined with tactile overlays, AI-generated navigation structures can be aligned with specific tactile markers, allowing users to interact with dynamically adjusted layouts while maintaining spatial consistency. For example, an AI-enhanced screen reader could group related elements—such as buttons, within a defined tactile region, enabling users to access application functions through direct touch instead of sequential scanning. This integration ensures that users retain a clear mental model of the interface, reinforcing the relationship between digital content and physical cues.

\subsection{Limitations}

Our findings demonstrate the potential of incorporating tactile overlays with screen readers to guide user interactions and convey spatial information but some limitations remain. \textit{Adaptive spatiotactile screen readers} may be less effective when (1) the density of actionable elements exceeds the overlay’s marker granularity (e.g., icon-rich drawing apps), and (2) when users cannot apply a physical overlay (public kiosks). 

Beyond these limitations,  we have yet to conduct tests with participants to validate how well our approach scales when content changes dynamically within a scenario. Future work should explore how to support applications that require frequent UI changes (e.g., live stock tickers with high frequency updates). Additionally, our one-hour study limited insights into long-term usability and user adaptation overtime. Our participant pool was also primarily Blind people, making our findings less generalizable to Low Vision users, who often require different overlay designs (as seen from our results). Finally, we recognize that certain interactions, such as requesting summaries with a four finger tap, would be more naturally executed through voice commands \cite{sharif2023understanding} rather than multi-finger gestures alone. Integrating voice input with touch-based spatial navigation could scaffold more intuitive and efficient interactions for \textit{adaptive spatiotactile screen readers}. 

Exploring these aspects and including participants design recommendations would enhance the broader applicability of \textit{adaptive spatiotactile screen readers} to create better screen reader experiences in touchscreen devices for BLV people.

%% file: src/09_conclusion.tex
\section{Conclusion}

In this paper, we introduced \DataTap{}, a prototype system that integrates tactile overlays with exploration-by-touch screen reader interactions, enabling BLV users to engage with content on touchscreen devices spatially (i.e., an \textit{adaptive spatiotactile screen reader)}.
We designed and iterated our prototype with formative usability and focus group studies and evaluated it through a user study with 12 BLV participants. 
Our findings showed the potential of this approach for enhancing self-driven exploration of screen content on touchscreen devices for BLV people. 
Through our discussion, we examined key considerations for building \textit{adaptive spatiotactile screen readers}, from the design questions that should guide the creation of tactile overlays, to future directions for expanding the use cases of \textit{adaptive spatiotactile screen readers} and extending the interactive expressiveness of this technique for touchscreen interfaces. 

Our work provides a foundation for future research exploring the use of passive or active tactile cues in combination with touch-context-aware screen reader prompts to increase the accessibility of touchscreen devices. We believe this technique affords novel, and powerful ways of expressing user intent on touchscreen devices that can significantly enhance BLV users ability to understand and engage with digital content.